\documentclass[aps,prb,twocolumn]{revtex4-1}
\usepackage{graphicx}
\usepackage[version=3]{mhchem}
\usepackage{booktabs}
\begin{document}

\title{Dangling bonds on the Cl- and Br-terminated Si(100) surfaces}

\author{T. V. Pavlova$^{1,2}$}
 \email{pavlova@kapella.gpi.ru}
\author{V. M. Shevlyuga$^{1}$}
\author{B. V. Andryushechkin$^{1}$}
\author{K. N. Eltsov$^{1}$}
\affiliation{$^{1}$Prokhorov General Physics Institute of the Russian Academy of Sciences, Moscow, Russia}
\affiliation{$^{2}$HSE University, Moscow, Russia}

\begin{abstract}

Halogen monolayer on a silicon surface is attracting active attention for applications in electronic device fabrication with individual impurities. To create a halogen mask for the impurities incorporation, it is desirable to be able to remove a single halogen atom from the surface. We report the desorption of individual halogen atoms from the Si(100)-2$\times$1-Cl and -Br surfaces in a scanning tunneling microscope (STM). Silicon dangling bonds (DBs) formed on the Si surface after halogen desorption were investigated using STM and the density functional theory. Three charge states: positive, neutral, and negative were identified. Our results show that the charge states of DBs can be manipulated, which will allow to locally tune the reactivity of the Cl- and Br-terminated Si(100) surfaces.

\end{abstract}

\maketitle

\section{Introduction}

Silicon dangling bonds (DBs) on an otherwise passivated Si(100)-2$\times$1 surface are known to have localized surface states in the band gap, which can be manipulated in a scanning tunneling microscope (STM) \cite{2007Raza, 2013Bellec, 2013Ye, 2015Labidi, 2016Kawai, 2017Scherpelz}. An unsaturated dangling bond of a Si atom is neutral (DB$^0$), while it will be referred to as positively charged (DB$^+$) if an electron is removed from the Si atom and negatively charged (DB$^-$) if an electron is added. The controlled change of the DB charge state allows to study interesting and complex physical phenomena, making the DBs attractive from the fundamental point of view \cite{1989Lyo, 2013Schofield, 2016Rashidi, 2018Wyrick}. From the practical point of view, DBs are promising for many applications in atomic scale silicon-based electronic devices due to the well-developed process of their precise creation on Si(100)-2$\times$1-H \cite{2009Haider, 2019He, 2020Achal, 2020Stock}.

Silicon dangling bonds affect the local reactivity of the surface, being active sites for such molecules as phosphine \cite{2003Schofield}, arsine \cite{2020Stock}, hydrocarbons \cite{2011Zikovsky}, HCl \cite{2011Li}, and I$_2$ \cite{2012Ferng}. The other study showed that ethylene adsorption on the Si(100)-H surface with DBs proceeds along a different pathway than on a clean Si(100) surface \cite{2009Mette}. Additionally, the selectivity of the surface reaction can be affected by the local charge on the surface, created by the DB charge. A negatively (positively) charged DB is nucleophilic (electrophilic), which can promote the adsorption of molecules with an empty lone pair of electrons (with a lone pair of electrons). For example, adsorption of BF$_3$ on negatively charged DBs due to preadsorption of trimethylamine (TMA) results in the formation of a donor-acceptor complex TMA-Si-Si-BF$_3$ \cite{2002Cao}. In addition, it was shown that the DB charge affects the self-assembly of organic nanostructures \cite{2012Ryan, 2014Piva}. Thus, the creation of single vacancies and the controlled switching of their charge is of great practical importance for site-selective surface reactions.

DBs on the H-terminated Si(100) surface have definitely received more attention than silicon DBs on the other surfaces. Taking into account that the interaction of Si(100) with chlorine is one of the key processes in the manufacture of microelectronic devices, the manipulation of DBs on a halogenated Si surface can be very promising for tuning surface reactions through site-selective reactivity. By studying the DBs on the silicon surface terminated by halogens (Hal = Cl, Br), we make a step toward the DBs applications on the technologically relevant Si(100)-2$\times$1-Hal surface. Recently, the Si(100)-2$\times$1-Hal surface has attracted research interest as a potential platform for the development of quantum devices with single impurities \cite{2018Pavlova, 2019Dwyer, 2020Pavlova, 2021Pavlova, 2021Frederick}. As well as hydrogen, halogens can be utilized as a resist on a silicon surface \cite{2021Pavlova, 2021Frederick} and, in addition, halogen resist is compatible with the Hal-containing dopants \cite{2019Dwyer}. Previously, it was shown that a strong Si-Cl bond allows desorption of Cl atoms together with Si atoms \cite{2020Pavlova} with the further goal of creating individual SiCl vacancies \cite{2018Pavlova}. In order to modify the halogenated Si(100) surface with atomic precision, the creation of single DBs is highly desirable.

DBs are naturally present on the Hal-terminated Si(100) surface at a Si surface atom when a terminating atom is missing, and they can be created \textit{in situ} in an STM. Although the Si(100)-2$\times$1-Hal surface containing halogen vacancies has been investigated in STM in several works \cite{2019Dwyer, 1993Boland, 1997Dohnalek, 1998Lyubinetsky, 2002Nakayama, 2003Nakamura}, the study of local desorption of halogen remains limited to a few examples. In the case of the Si(100)-2$\times$1-Cl surface, Dohnalek et al. reported desorption of pairs of Cl atoms under combined action of an STM tip and a laser pulse \cite{1997Dohnalek}, whereas Dwyer et al. \cite{2019Dwyer} demonstrated multiple Cl vacancies creation in STM. Single vacancies were produced only on the (111) face of the chlorinated silicon surface \cite{2001Nakamura, 2007Nakamura}. In the case of the Br-terminated Si(111) surface, Br atoms were removed along with Si adatoms at the atomic-layer etching in STM \cite{1999Mochiji, 2000Mochiji}.

An additional motivation for this work is to study charge states of the DB on the halogenated Si(100) surface. In previous works, vacancies were visualized as depressions or protrusions. Specifically, depressions in STM images at both polarities of bias voltage were identified as single Cl \cite{2019Dwyer, 1998Lyubinetsky} or Br \cite{2002Nakayama} vacancies on the Si(100)-2$\times$1-Hal surface, in agreement with simulated STM images \cite{2004Lee}. In Refs.~\cite{1993Boland, 2003Nakamura}, protrusions were attributed to DBs at Cl vacancy sites in STM images. In the case of the H-terminated Si(100) surface, it is well known that different visualization is determined by the charge localized on DB \cite{2013Bellec, 2015Labidi, 2016Kawai}. However, the charge states of DBs on Si(100)-2$\times$1-Hal and their influence on the visualization of vacancies have yet to be investigated.

This work is aimed to create single vacancies and study the charge states of DBs on the Cl- and Br-terminated Si(100)-2$\times$1 surfaces. Single Cl  and Br vacancies have been successfully created by applying voltage pulses in an STM. The identification of positive, neutral, and negative charge states of the DB was carried out using density functional theory (DFT). The electronic structure of DBs on the halogenated surface is discussed in comparison with that for a well-studied case of the hydrogenated surface. The DB charge was switched between positive and negative in a controlled manner by varying the positive voltage at the sample.

\section{Methods}

\subsection{Experimental methods}

The experiments were carried out in an ultra-high vacuum (UHV) with a base pressure of 5$\times$10$^{-11}$\,Torr. All measurements and atomic manipulations were performed using a low-temperature scanning tunneling microscope GPI CRYO (SigmaScan Ltd.) at 77\,K. We used B- (p-type, 1\,$\Omega$\,cm) and P-doped (n-type, 0.1\,$\Omega$\,cm) Si(100) samples. Si(100) samples were outgassed at 870\,K for several days in UHV and then cleaned by direct-current flash annealing at 1470\,K. Molecular chlorine and bromine were introduced with a partial pressure of 10$^{-8}$\,Torr during 100--200\,s at a sample temperature of 370--420\,K after the flash heating was switched off. We used both electrochemically etched polycrystalline W tips and mechanically cut Pt-Rh tips for scanning and manipulation experiments. All STM images have been processed with the WSXM software \cite{WSXM}.

\subsection{Computational methods}

The spin-polarized DFT calculations were performed using the projector augmented wave method implemented in the Vienna \textit{ab initio} simulation package (VASP) \cite{1996Kresse, 1999Kresse}. For the exchange-correlation potential, generalized gradient approximation (GGA) of Perdew-Burke-Ernzerhof (PBE) \cite{1996Perdew} was employed. The van der Waals correction was included by using the DFT-D2 method developed by Grimme \cite{2006Grimme}. The kinetic-energy cutoff was set to 400\,eV. The Si(100)-2$\times$1 surface was modeled by a sixteen layer slab with a 6$\times$6 supercell. A vacuum spacing of 15\,{\AA} thickness was used to avoid interactions between the neighboring surfaces.
Calculations of the electronic structure were carried out for DBs on chlorinated, brominated, and hydrogenated surfaces. The adsorbate atoms were placed on the upper surface to form a Si(100)-2$\times$1-Cl, -Br, and -H structures, whereas the bottom Si surface was saturated with hydrogens. The bottom two Si layers were kept in their bulk positions, while the coordinates of other atoms are fully relaxed until the residual forces on each atom are equal to or less than 0.01\,eV/\,{\AA}. To simulate positive or negative charge states, the electron was removed or added to the supercell, respectively. The Brillouin zone was sampled by a 4$\times$4$\times$1 k-point grid. For the electronic density of states (DOS) calculations, the reciprocal cell was integrated using a 8$\times$8$\times$1 k-point grid. STM images were calculated within the Tersoff-Hamann approximation \cite{1985Tersoff}. The voltage in simulated STM images is indicated relative to the valence-band maximum (VBM).

\section{Results and discussion}

\subsection{Creation of DBs}

The Si(100)-2$\times$1-Hal surface consists of Si-dimer rows, and each Si atom of the dimer is terminated by a halogen atom. The Br-terminated surface is very similar in appearance to the Cl-terminated one. For halogen extraction, the tip was positioned over a certain desired lateral position of the surface, the feedback loop was disabled, and a rectangular voltage pulse was applied to the sample. Before the pulse, the tip position above the Si(100)-2$\times$1-Hal surface remained the same as in the scanning mode. The voltage leading to the creation of single vacancies ranged from $\pm 2$ to $\pm 5$ V,  and the pulse duration varied from 1 to 100 ms. We used $\pm 2$ V voltage pulses as starting values since they very rarely caused desorption, so they can be considered threshold voltages.  We raised the voltage until one halogen atom was desorbed and then continued to use this pulse. If the pulse no longer led to desorption, the voltage was increased by 0.1 V. When the total increase in voltage to about 1 V did not lead to desorption, the pulse duration was increased. These parameters depended on the tip state, which changed after the pulse, as well as on the local electrostatic environment of the selected surface area, which is determined by nearby impurities and defects. The presence or absence of desorption event was detected by imaging the surface immediately after the pulse was applied.

\begin{figure}[h]
\begin{center}
 \includegraphics[width=\linewidth]{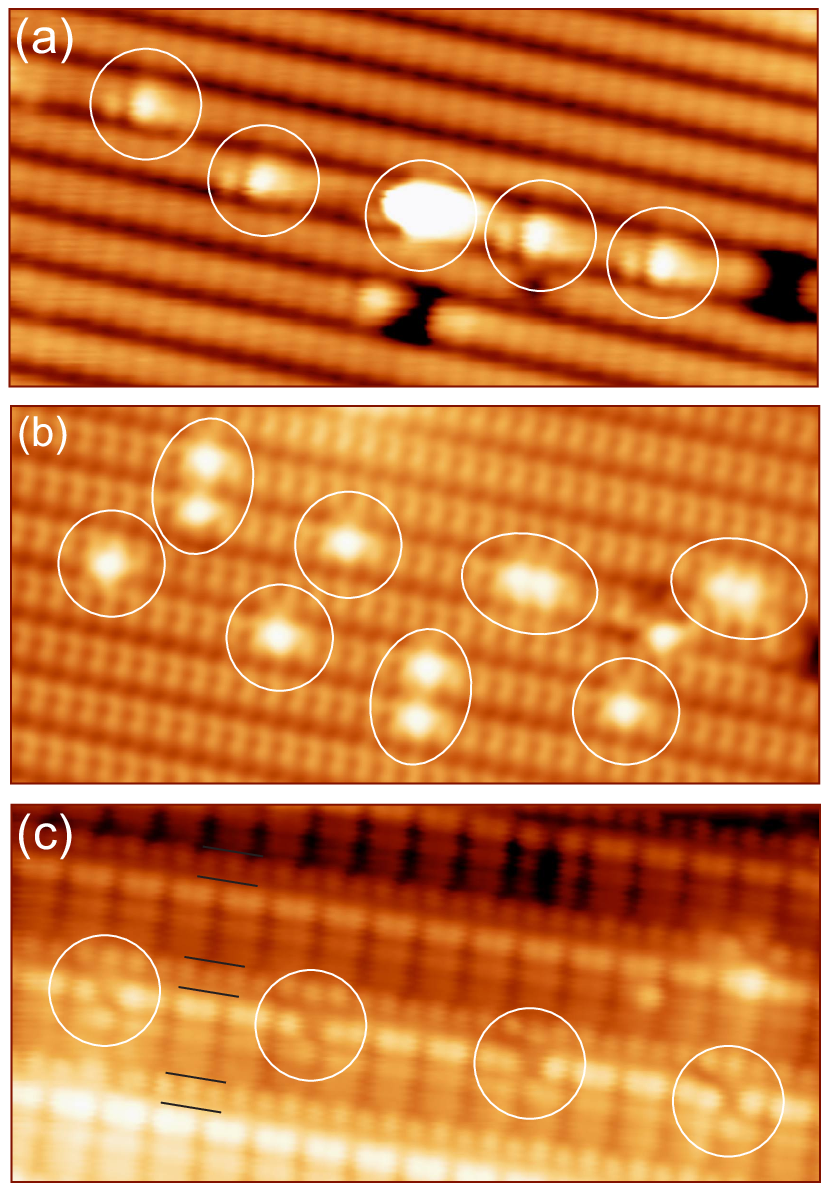}
\caption{\label{Cr_neg} Successive halogen extraction from the Si(100)-2$\times$1-Cl (a, b) and Si(100)-2$\times$1-Br (c) surfaces by a negative voltage pulse. Desorption events that occurred at a single pulse are united by a circle. (a, b) Filled state STM images (12.1$\times$5.7\,nm$^2$, $U_s =-2.5$\,V, I$_t$ = 0.4\,nA, B-doped Si, W tip) of Cl vacancies created by pulses of $-4.0$\,V, 0.4\,nA, 1\,ms. In (a), the third vacancy has changed during the scanning process. (c) Empty state STM image (12.1$\times$5.7\,nm$^2$, $U_s =+2.2$\,V, I$_t$ = 2.0\,nA, P-doped Si, Pt-Rh tip) of four Br vacancies on the step edge created by pulses of $-2.2$\,V, 3\,ms. Step edges are marked with black lines.}
\end{center}
\end{figure}

Figure~\ref{Cr_neg} shows vacancies created by a negative voltage pulse on the Cl- and Br-passivated Si(100) surfaces. We created Br vacancies at the edge of an atomic step at a distance of 7--8 atoms from each other. The DBs on the chlorinated B-doped Si surface look like protrusions extended over almost the entire dimer and neighboring atoms (Fig.~\ref{Cr_neg}a,b), which is consistent with our calculations for a negatively charged DB (as we will show below). Instead, the DBs on the brominated P-doped Si surface (Fig.~\ref{Cr_neg}c) are neutral. An STM tip removes atoms from the surface more efficiently if it is not atomically sharp \cite{2017Moller}, which is the reason for the lack of good resolution in the STM image recorded immediately after the vacancies creation.

\begin{figure*}[t!]
\begin{center}
 \includegraphics[width=\linewidth]{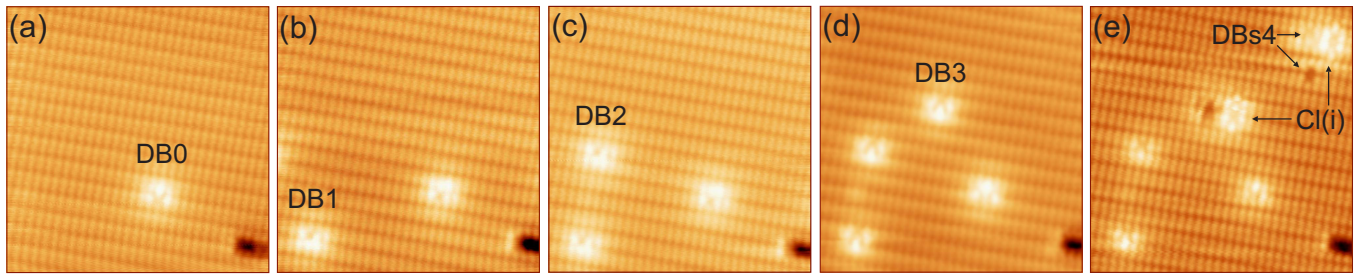}
\caption{\label{Cr_pos} Sequentially recorded empty state STM images (12$\times$12\,nm$^2$, $U_s =+2.5$\,V, I$_t$ = 2.0\,nA, B-doped Si, Pt-Rh tip) of Cl extraction from the Si(100)-2$\times$1-Cl surface by positive voltage pulses of $+5$\,V, 0.4\,nA, 100\,ms. (a) Initial surface with DB0; (b--d) DB1, DB2, and DB3 created by three single pulses; (e) two DBs (DBs4) and two Cl(i) formed as a result of the fourth pulse.}
\end{center}
\end{figure*}

By applying positive voltage pulses, we were able to sequentially create several single DBs. The vacancies created on B-doped Si(100)-2$\times$1-Cl are shown in Fig.~\ref{Cr_pos}, and those created on P-doped are shown below (see Fig.~\ref{fig5}). Usually, the next pulse led to the simultaneous desorption of multiple atoms or/and halogen insertion into the Si(100)-2$\times$1-Hal surface. Initially, there was a bright object on the surface, DB0 (Fig.~\ref{Cr_pos}a), around which three identical objects were successively created (Fig.~\ref{Cr_pos}b--d). As will be shown below, these bright objects are positively charged dangling bonds (DB0--DB3). After the fourth pulse, two Cl atoms were extracted and two Cl atoms were inserted into the surface (Fig.~\ref{Cr_pos}e). The inserted chlorine atoms, Cl(i), are located in the bridge-bonded positions inside a Cl saturated Si dimer \cite{2020PavlovaPRB}. After the appearance of Cl(i) on the surface, DB3 visually changed. This can be explained by the fact that positively charged Cl(i) induces local band bending, which affects the visualization of the nearby DB. Note that the insertion of adatoms into the surface also occurs during the STM lithography on Si(100)-2$\times$1-H \cite{2014Ballard2}, and they can occupy the same adsorption sites on the surface \cite{2020PavlovaPCCP} as in the case of chlorine.

In contrast to the halogen desorption, the hydrogen desorption from the Si(100)-2$\times$1-H surface has been developed and optimized in a large number of works and now it is realized with the real-atomic precision \cite{2003Schofield, 2017Moller, 2018Achal, 2018Randall}. Two regimes of H desorption are commonly distinguished: at low ($<$3.5 V) and high voltages ($>$6 V) \cite{1995Shen}. Atomic accuracy is achieved at low voltage, while at high voltage, desorption of several atoms occurs at once, which makes it possible to clean large areas from hydrogen. In the low-energy regime, the desorption mechanism is a vibrational heating of a Si-H bond, leading to bond cleavage  \cite{1995Shen}. Indeed, the H-Si vibration can be excited several times before it relaxes due to the extra long lifetime of the stretching mode. Since the desorption yield follows a power law as a function of the tunnel current with a power equal to the number of excitations, the multi-electron vibrational heating mechanism of hydrogen desorption has a higher yield in the high-current regime. On the contrary, the Si-Hal vibrations are strongly coupled to the silicon phonon modes; therefore, multiple excitations of Cl and Br vibrations by an STM pulse are unlikely \cite{2020Pavlova}. Consequently, such an efficient vibrational mechanism of H desorption does not work for Cl and Br. This suggests that the halogen desorption yield at any current will be lower than the hydrogen desorption yield in the high-current regime.

We assume that the mechanism of halogen desorption in the low-energy regime also most likely occurs due to the electronic excitation of the Si-adsorbate bond, but by only one carrier (at $U_s>0$, an electron is captured into the Si-Hal antibonding orbital, and at $U_s<0$, a hole is captured into the Si-Hal bonding orbital). The capture of one electron (hole) in the Si-adsorbate orbital is also responsible for the diffusion of halogens \cite{2002Nakayama, 2003Nakamura} and hydrogen \cite{2010Bellec} over the Si(100) surface. However, in the diffusion mechanism, the carrier transfer to the Si-adsorbate orbital can occur through delocalized Si surface states, making the diffusion process nonlocal. Unlike diffusion, hydrogen desorption is local, which is confirmed by the creation of single vacancies with atomic precision. Our observation that halogens are desorbed directly under the tip (or from nearby dimers if the tip apex is not atomically sharp) also suggests that desorption is a local process.

At a pulse with a voltage slightly higher (3--4 V) than that for single atom desorption (2--4 V), a pair of H atoms desorbs \cite{2006Tong}. Interestingly, two atoms are desorbed from adjacent dimers rather than from one dimer. We also sometimes observed pairwise desorption of halogen atoms, always from neighbouring dimers along or across the dimer row, and not from the same dimer (Fig.~\ref{Cr_neg}b). The Hal$_2$ desorption from neighbouring dimers indicates that the mechanism of recombinative desorption can be similar to that for H$_2$.

We did not notice any difference in the desorption of single halogen atoms from a step edge and a terrace. This can be explained by the fact that the halogen atom is bonded to one Si atom of the dimer both on the step and on the terrace. However, the Si-Br bond (3.59 eV) is weaker than the Si-Cl bond (4.27 eV) \cite{2004Lee}, and therefore the voltage at which Br desorption occurs should be less than for Cl. In our work, we were unable to detect a difference in the extraction of Br and Cl since the voltage pulse variations at the creation of vacancies exceeded the difference between the Br-Cl and Si-Cl bonds (0.7 eV). We attribute the voltage pulse variations to two factors: different tip states and non-uniform impurity distribution, which, as will be shown below, can lead to a 2 V difference in charge switching of DBs located at a distance of several nanometers from each other.

Although we have demonstrated the removal of individual Cl and Br atoms, we did not achieve halogen desorption with atomic precision. However, we believe that it is possible to create halogen vacancies with atomic precision since the mechanism of halogen desorption seems to be local. The arising spurious desorption events in the form of undesirable vacancies can, in principle, be eliminated by capping the vacancy with a Hal atom, as was done in the case of the H-terminated Si(100) surface \cite{2018Achal, 2017Niko}. Since halogen injection from a tip into a surface has been demonstrated \cite{2020PavlovaPRB}, such a process could be used to correct erroneously created vacancies. To further improve accuracy and control of desorption, a very stable tip, uniform doping in the near-surface region, and the use of an STM feedback-tracking protocol to detect a desorption event \cite{2017Moller} are required.

\subsection{Charge states of DBs}

On the H-terminated Si(100)-2$\times$1 surface, DBs are well studied \cite{2007Raza, 2013Bellec, 2013Ye, 2015Labidi, 2016Kawai, 2017Scherpelz}. Chlorine, bromine, and hydrogen form the same structure 2$\times$1 on the Si(100) surface, but the electronegativity of chlorine and bromine is higher than that of hydrogen. The question that we seek to answer in this section: are the geometry and electronic structure of the DB on the Si(100)-2$\times$1-Hal surface similar to those on the well-studied Si(100)-2$\times$1-H surface? To address this question, we performed DFT calculations of surface structures and DOS of DBs for the Cl-, Br-, and H-terminated Si(100)-2$\times$1 surfaces.

The higher electronegativity of chlorine causes a higher polarization of the Si-Cl bond compared to Si-H. As a result, the calculated work function of the Si(100)-2$\times$1-Cl surface is about 1.0 eV higher than that for Si(100)-2$\times$1-H. Similar values were obtained for the Si(111)-1$\times$1-Cl surface both theoretically (1.55 eV) \cite{2006Blomquist} and experimentally (1.2--1.5 eV) \cite{2005Lopinski}. However, the shift of the electrostatic potential from the vacuum level does not lead to the appearance of band bending, as was shown for the Si(111)-1$\times$1-Cl surface \cite{2006Blomquist}. Therefore, electronegative chlorine (and bromine) do not induce band banding, leaving the bands flat, as in the case of hydrogen on the Si(100) surface.

\begin{figure}[h!]
 \begin{center}
 \includegraphics[width=\linewidth]{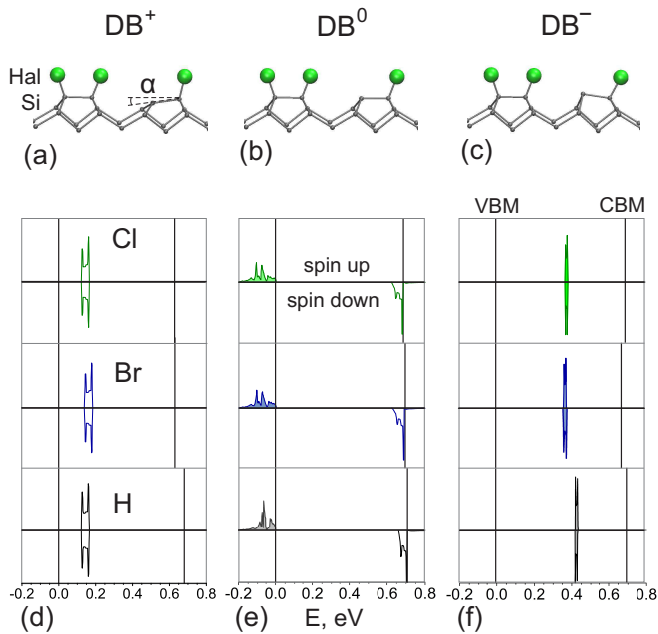}
\caption{\label{figDBs1} Surface structures and DOS of DBs in three different charge states on the Cl-, Br-, and H-terminated Si(100)-2$\times$1 surfaces. Side view of the surface structures with DBs are shown for positive (a), neutral (b), and negative (c) charge states. Halogen atoms are marked in green, Si atoms in gray. Spin-resolved DOS of the Si atom holding the DB are plotted relative to the VBM for positive (d), neutral (e), and negative (f) charge states. For the Cl-, Br-, and H-terminated surfaces, the DOS and CBM are shown with green, blue, and black lines, respectively. The VBM and CBM are plotted for Si atoms in two layers in the middle of the slab. The CBM is slightly different for all adsorbates since the adsorbate atoms slightly affect the electronic states of the Si atoms in the middle of the slab.}
\end{center}
\end{figure}

Surface structures with the DB$^+$, DB$^0$, and DB$^-$ differ in the tilt angle ($\alpha$) of the Si dimer relative to the horizontal plane (Fig.~\ref{figDBs1}a--c). Downward-tilted configuration of the surface structure with the DB$^+$ indicates sp$^2$-like hybridization of the orbitals, while in the case of the DB$^0$ and DB$^-$ the hybridization changes toward sp$^3$. The tilt angles of the Si dimer holding the DB$^+$, DB$^0$, and DB$^-$ on the Si(100)-2$\times$1-Hal surface are very close to those in the case of Si(100)-2$\times$1-H (Table \ref{table1}), which in turn is consistent with previous work \cite{2016Kawai}. Therefore, the geometry of surface structures with DBs on Cl- and Br-terminated Si(100)-2$\times$1 turned out to be approximately the same as for the H-terminated Si(100)-2$\times$1 surface.

\begin{table}[h!]
\renewcommand{\tabcolsep}{0.3cm}
\begin{center}
\caption{The tilt angle ($\alpha$) of the dimer with the DB on Cl-, Br-, and H-terminated Si(100)-2$\times$1 surfaces. The results of Ref.~\cite{2016Kawai} are given in the last line.}
    \label{table1}
    \begin{tabular}{cccc}
    \toprule
   Adsorbate &  DB$^+$ & DB$^0$ & DB$^-$ \\ \hline \hline
    \midrule
    Cl & $-9.4^{\circ}$ & $-0.7^{\circ}$ & $+8.0^{\circ}$  \\ \hline
    Br & $-9.5^{\circ}$ & $-1.0^{\circ}$ &  $+7.7^{\circ}$  \\ \hline
    H & $-9.2^{\circ}$ & $-0.9^{\circ}$ & $+7.3^{\circ}$  \\ \hline
    H\cite{2016Kawai} & $-10.2^{\circ}$ & $-1.2^{\circ}$ & $+7.6^{\circ}$  \\
       \bottomrule
\end{tabular}
\end{center}
\end{table}

Figures~\ref{figDBs1}d--f show electron levels of a Si atom holding the DB on the Si(100)-2$\times$1 surfaces passivated with Cl, Br, and H. A doubly unoccupied electron level of the DB$^+$ lies inside the band gap (Fig.~\ref{figDBs1}d). The DB$^0$ has an occupied level below the VBM and an unoccupied level in the band gap just below the conduction-band minimum (CBM) (Fig.~\ref{figDBs1}e). A doubly occupied electron level of the DB$^-$ lies inside the band gap (Fig.~\ref{figDBs1}f), about 0.25 eV higher than the DB$^+$ level. According to our calculations, the DOS of the DB$^+$, DB$^0$, and DB$^-$ are almost the same for chlorinated, brominated, and hydrogenated Si(100) surfaces (Table \ref{table2}). This was expected, since the DB orbital is much more hybridized with orbitals of three neighboring Si atoms than with orbitals of adsorbate atoms. Comparing the DB levels with previous calculations on the hydrogenated surface \cite{2016Kawai, 2017Scherpelz, 2013Schofield}, we found that they are in qualitative agreement. Note that for the (111) silicon face, the calculated DB$^0$ levels on the chlorinated surface also agree with those on the hydrogenated surface \cite{2006Blomquist}.

\begin{table}[h!]
\renewcommand{\tabcolsep}{0.3cm}
\begin{center}
\caption{Positions of energy levels of DBs on Cl-, Br-, and H-terminated Si(100)-2$\times$1 surfaces. All energies are given in electronvolts with respect to the VBM. The energy levels of previous works \cite{2016Kawai, 2017Scherpelz, 2013Schofield} are given at the $\Gamma$-point.}
    \label{table2}
    \begin{tabular}{cccc}
        \toprule
     Adsorbate &  DB$^+$ & DB$^0$ & DB$^-$ \\ \hline \hline
    \midrule
  Cl & 0.15 & $-0.13$; 0.67 & 0.38   \\ \hline
  Br & 0.16 & $-0.13$; 0.67 & 0.37   \\ \hline
  H & 0.15 & $-0.10$; 0.70 & 0.42   \\ \hline
  H\cite{2016Kawai} & 0.2 &  --- & 0.4   \\ \hline
  H\cite{2017Scherpelz}  & --- &  $-0.1$; 0.7 & ---   \\ \hline
  H\cite{2013Schofield} & 0.01 &  0.01; 0.56 & 0.35   \\
    \bottomrule
\end{tabular}
\end{center}
\end{table}

\begin{figure}[h!]
 \begin{center}
 \includegraphics[width=\linewidth]{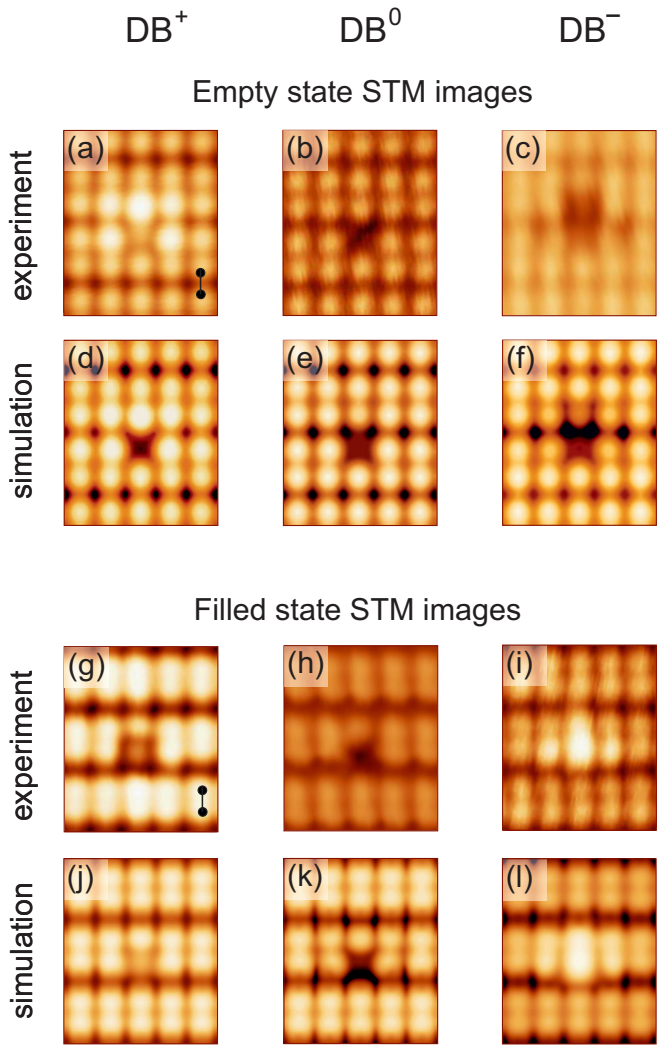}
\caption{\label{fig4} Empty (a--f) and filled (g--l) state STM images of the DB$^+$, DB$^0$, and DB$^-$ on the Si(100)-2$\times$1-Cl surface. Experimental images were obtained at I$_t$ = 1.0\,nA; $U_s =+2.5$\,V (a), $+2.4$\,V (b), $+2.3$\,V (c), $-2.8$\,V (g), $-2.2$\,V (h), $-3.5$\,V (i). Theoretical images were calculated at $U =+2.0$\,V (d--f) and $-1.0$\,V (j--l). Silicon dimers are marked by dumbbells in (a) and (g). The voltage used to calculate STM images may differ from the voltage in the experiment since the calculated band gap is 0.4 eV less than the experimental one. In addition, randomly distributed impurities induce band bending, which is not taken into account in the calculations.}
\end{center}
\end{figure}

Figure~\ref{fig4} shows STM images of the DB$^+$, DB$^0$, and DB$^-$ on the Si(100)-2$\times$1-Cl surface. To identify the charge states of DBs in experimental STM images, we simulate STM images of the DB in three charge states. In empty state STM images, three atoms close to the DB$^+$ look brighter (Fig.~\ref{fig4}a,d), whereas they look darker in the case of the DB$^-$ (Fig.~\ref{fig4}c,f). The reason for the brighter (darker) Cl atoms near the DB$^+$ (DB$^-$) in empty state STM images is the hybridization of the Cl orbitals with the unoccupied (occupied) DB orbitals. In filled state STM images, the doubly occupied DB$^-$ looks bright (Fig.~\ref{fig4}i,l). At low voltage, the DB$^-$ has a characteristic dark halo as in the case of a hydrogenated surface \cite{2015Labidi, 2013Schofield}. The DBs on the Br-terminated surface have the same visualization in STM images as on the Cl-terminated surface due to the similarity of the energy spectrum.

The charge state of the DB depends on the position of the DB$^+$ and DB$^-$ levels relative to the Fermi level. The DB$^+$ and DB$^-$ are thermodynamically preferable if their levels are located above and below the Fermi level, respectively. However, we have observed all charge states of DBs shown in Fig.~\ref{fig4} on both p-type (1\,$\Omega$\,cm) and n-type (0.1\,$\Omega$\,cm) Si(100) samples. The ability to use samples with different doping levels is explained by the fact that all samples are more intrinsic in the near-surface region of about tens of nanometers. This feature originates from dopants depletion near the surface due to flash annealing of the samples to 1470\,K \cite{2012Pitters}. Note that when we imaged the DBs after retracting the tip, the DBs retained the same charge state; therefore, the charge state of the DBs is stable (at least for several hours). Very rarely, we have observed a spontaneous change in the DBs charge during scanning without a change in voltage.

The DB charge can be switched by changing the voltage in the STM. In Ref. \cite{2013Bellec}, the switching of the DB charge with a change in voltage was explained by a shift in the positions of the DB levels relative to the Fermi level due to the tip induced band bending (TIBB). In our case, a positive voltage on the sample (a negative voltage on the tip) leads to upward band bending, so that at a sufficiently high voltage, the levels DB$^+$, DB$^0$, and DB$^-$ turn out to be above the Fermi level. Consequently, the DB$^+$ turns out to be the most stable, and all DBs become positively charged. As the voltage decreases, the band bending decreases, and, accordingly, the DB levels shift downward. When the DB$^-$ turns out to be below the Fermi level, DB becomes negatively charged. Note that the DB charge can also be switched by changing the voltage polarity during scanning, as was demonstrated for DBs on the Si(100)-2$\times$1-H surface \cite{2013Bellec}.

\begin{figure*}[t!]
\begin{center}
 \includegraphics[width=0.9\linewidth]{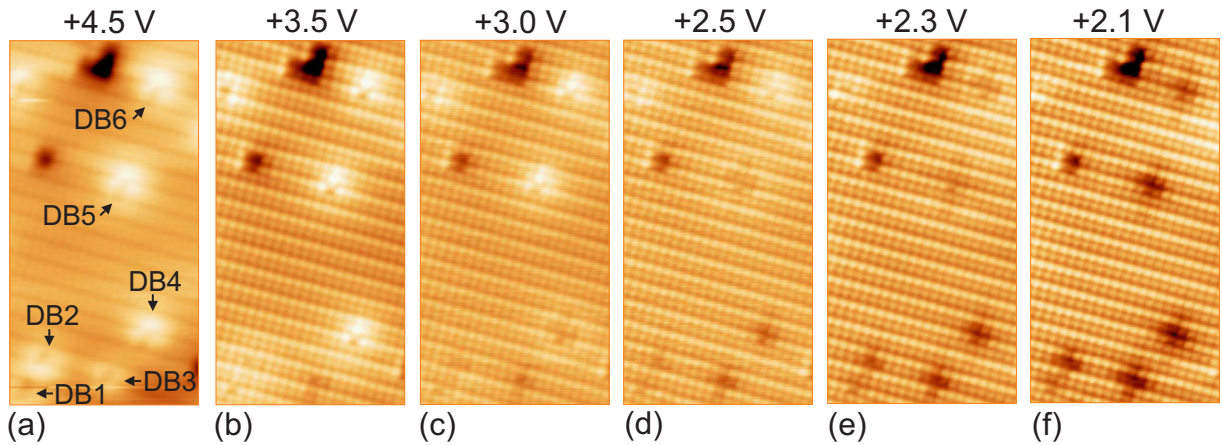}
\caption{\label{fig5} Controllable switching of the charge state from positive to negative for six DBs created by positive voltage pulses of $+4.4$\,V, 1.0\,nA, 3\,ms on the Si(100)-2$\times$1-Cl surface. Sequentially recorded empty state STM images (7.4$\times$14.2\,nm$^2$, I$_t$ = 1.0\,nA, $ U_s$ is indicated above each image, P-doped Si, Pt-Rh tip) of DBs on Si(100)-2$\times$1-Cl. The bright and dark halos around the DBs correspond to the positively and negatively charged DBs, respectively. (a) At $+4.5$ V, all DBs are positively charged. (b--e) With decreasing voltage, conversion to negative charge state gradually occurs. (f) At $+2.1$ V, all DBs are negatively charged. The strong variation in the charging voltage (by about 2 V) is due to the highly inhomogeneous electrostatic potential of the near-surface region.}
\end{center}
\end{figure*}

In our experiments, the voltage required to switch the charge states varied strongly for different DBs. More specifically, the variations in the voltage were observed not only for DBs on different samples but also for DBs in different places on the same STM image. Figure~\ref{fig5} shows variation in the voltage at which charging of DBs occurs. At high voltage, all the DBs are positively charged (Fig.~\ref{fig5}a), but as the voltage decreases, the DBs$^+$ at the bottom of the STM images start to switch to the DBs$^-$ (Fig.~\ref{fig5}b--e), and finally, all the DBs become negatively charged (Fig.~\ref{fig5}f). Such a wide spread of voltages is explained by the inhomogeneous electrostatic landscape of the surface due to the non-uniform distribution of subsurface dopants \cite{2015Labidi}. Thus, the randomly distributed subsurface dopants drastically affect the voltage at which the charging of DBs occurs and can also affect the pulse voltage at which atoms are desorbed from the surface.

\section{Conclusions}

In conclusion, we have demonstrated the creation of a series of individual vacancies on the Si(100)-2$\times$1-Cl and Si(100)-2$\times$1-Br surfaces by voltage pulses in an STM. On the Cl-, Br-, and H-terminated surfaces, the calculated geometry and electronic structure of DBs turned out to be similar, and, therefore, DBs should demonstrate similar electronic properties and charging behavior. The DBs were charged in a controlled manner, and it was shown that the voltage required to switch the charge of the DB depends on the local electrostatic environment. The controlled change in the DBs charge state will make it possible to study interesting physical phenomena on the halogenated Si(100) surface, such as negative differential resistance \cite{1989Lyo, 2016Rashidi}, simulation of artificial molecules \cite{2013Schofield, 2018Wyrick}, etc., previously observed only on the hydrogenated surface. Also, the demonstrated manipulation of the DB charge can be used to locally tune the reactivity of a halogenated surface. To create Cl and Br vacancies with atomic precision desired for various technological applications on the halogenated Si(100) surface, further work is needed, in particular, to improve the stability of the STM tip.

\section*{Acknowledgments}
The reported study was funded by RFBR, project number 20-02-00783. We also thank the Joint Supercomputer Center of RAS for providing the computing power.

\bibliography{ApplSurfSci_Pavlova_DB_arxiv}

\begin{thebibliography}{56}%
\makeatletter
\providecommand \@ifxundefined [1]{%
 \@ifx{#1\undefined}
}%
\providecommand \@ifnum [1]{%
 \ifnum #1\expandafter \@firstoftwo
 \else \expandafter \@secondoftwo
 \fi
}%
\providecommand \@ifx [1]{%
 \ifx #1\expandafter \@firstoftwo
 \else \expandafter \@secondoftwo
 \fi
}%
\providecommand \natexlab [1]{#1}%
\providecommand \enquote  [1]{``#1''}%
\providecommand \bibnamefont  [1]{#1}%
\providecommand \bibfnamefont [1]{#1}%
\providecommand \citenamefont [1]{#1}%
\providecommand \href@noop [0]{\@secondoftwo}%
\providecommand \href [0]{\begingroup \@sanitize@url \@href}%
\providecommand \@href[1]{\@@startlink{#1}\@@href}%
\providecommand \@@href[1]{\endgroup#1\@@endlink}%
\providecommand \@sanitize@url [0]{\catcode `\\12\catcode `\$12\catcode
  `\&12\catcode `\#12\catcode `\^12\catcode `\_12\catcode `\%12\relax}%
\providecommand \@@startlink[1]{}%
\providecommand \@@endlink[0]{}%
\providecommand \url  [0]{\begingroup\@sanitize@url \@url }%
\providecommand \@url [1]{\endgroup\@href {#1}{\urlprefix }}%
\providecommand \urlprefix  [0]{URL }%
\providecommand \Eprint [0]{\href }%
\providecommand \doibase [0]{http://dx.doi.org/}%
\providecommand \selectlanguage [0]{\@gobble}%
\providecommand \bibinfo  [0]{\@secondoftwo}%
\providecommand \bibfield  [0]{\@secondoftwo}%
\providecommand \translation [1]{[#1]}%
\providecommand \BibitemOpen [0]{}%
\providecommand \bibitemStop [0]{}%
\providecommand \bibitemNoStop [0]{.\EOS\space}%
\providecommand \EOS [0]{\spacefactor3000\relax}%
\providecommand \BibitemShut  [1]{\csname bibitem#1\endcsname}%
\let\auto@bib@innerbib\@empty
\bibitem [{\citenamefont {Raza}(2007)}]{2007Raza}%
  \BibitemOpen
  \bibfield  {author} {\bibinfo {author} {\bibfnamefont {H.}~\bibnamefont
  {Raza}},\ }\href@noop {} {\bibfield  {journal} {\bibinfo  {journal} {Phys.
  Rev. B}\ }\textbf {\bibinfo {volume} {76}},\ \bibinfo {pages} {045308}
  (\bibinfo {year} {2007})}\BibitemShut {NoStop}%
\bibitem [{\citenamefont {Bellec}\ \emph {et~al.}(2013)\citenamefont {Bellec},
  \citenamefont {Chaput}, \citenamefont {Dujardin}, \citenamefont {Riedel},
  \citenamefont {Stauffer},\ and\ \citenamefont {Sonnet}}]{2013Bellec}%
  \BibitemOpen
  \bibfield  {author} {\bibinfo {author} {\bibfnamefont {A.}~\bibnamefont
  {Bellec}}, \bibinfo {author} {\bibfnamefont {L.}~\bibnamefont {Chaput}},
  \bibinfo {author} {\bibfnamefont {G.}~\bibnamefont {Dujardin}}, \bibinfo
  {author} {\bibfnamefont {D.}~\bibnamefont {Riedel}}, \bibinfo {author}
  {\bibfnamefont {L.}~\bibnamefont {Stauffer}}, \ and\ \bibinfo {author}
  {\bibfnamefont {P.}~\bibnamefont {Sonnet}},\ }\href@noop {} {\bibfield
  {journal} {\bibinfo  {journal} {Phys. Rev. B}\ }\textbf {\bibinfo {volume}
  {88}},\ \bibinfo {pages} {241406(R)} (\bibinfo {year} {2013})}\BibitemShut
  {NoStop}%
\bibitem [{\citenamefont {Ye}\ \emph {et~al.}(2013)\citenamefont {Ye},
  \citenamefont {Min}, \citenamefont {Martin}, \citenamefont {Rockett},
  \citenamefont {Aluru},\ and\ \citenamefont {Lyding}}]{2013Ye}%
  \BibitemOpen
  \bibfield  {author} {\bibinfo {author} {\bibfnamefont {W.}~\bibnamefont
  {Ye}}, \bibinfo {author} {\bibfnamefont {K.}~\bibnamefont {Min}}, \bibinfo
  {author} {\bibfnamefont {P.~P.}\ \bibnamefont {Martin}}, \bibinfo {author}
  {\bibfnamefont {A.~A.}\ \bibnamefont {Rockett}}, \bibinfo {author}
  {\bibfnamefont {N.}~\bibnamefont {Aluru}}, \ and\ \bibinfo {author}
  {\bibfnamefont {J.~W.}\ \bibnamefont {Lyding}},\ }\href@noop {} {\bibfield
  {journal} {\bibinfo  {journal} {Surf. Sci.}\ }\textbf {\bibinfo {volume}
  {609}},\ \bibinfo {pages} {147} (\bibinfo {year} {2013})}\BibitemShut
  {NoStop}%
\bibitem [{\citenamefont {Labidi}\ \emph {et~al.}(2015)\citenamefont {Labidi},
  \citenamefont {Taucer}, \citenamefont {Rashidi}, \citenamefont {Koleini},
  \citenamefont {Livadaru}, \citenamefont {Pitters}, \citenamefont {Cloutier},
  \citenamefont {Salomons},\ and\ \citenamefont {Wolkow}}]{2015Labidi}%
  \BibitemOpen
  \bibfield  {author} {\bibinfo {author} {\bibfnamefont {H.}~\bibnamefont
  {Labidi}}, \bibinfo {author} {\bibfnamefont {M.}~\bibnamefont {Taucer}},
  \bibinfo {author} {\bibfnamefont {M.}~\bibnamefont {Rashidi}}, \bibinfo
  {author} {\bibfnamefont {M.}~\bibnamefont {Koleini}}, \bibinfo {author}
  {\bibfnamefont {L.}~\bibnamefont {Livadaru}}, \bibinfo {author}
  {\bibfnamefont {J.}~\bibnamefont {Pitters}}, \bibinfo {author} {\bibfnamefont
  {M.}~\bibnamefont {Cloutier}}, \bibinfo {author} {\bibfnamefont
  {M.}~\bibnamefont {Salomons}}, \ and\ \bibinfo {author} {\bibfnamefont
  {R.~A.}\ \bibnamefont {Wolkow}},\ }\href@noop {} {\bibfield  {journal}
  {\bibinfo  {journal} {New J. Phys.}\ }\textbf {\bibinfo {volume} {17}},\
  \bibinfo {pages} {073023} (\bibinfo {year} {2015})}\BibitemShut {NoStop}%
\bibitem [{\citenamefont {Kawai}\ \emph {et~al.}(2016)\citenamefont {Kawai},
  \citenamefont {Neucheva}, \citenamefont {Yap}, \citenamefont {Joachim},\ and\
  \citenamefont {Saeys}}]{2016Kawai}%
  \BibitemOpen
  \bibfield  {author} {\bibinfo {author} {\bibfnamefont {H.}~\bibnamefont
  {Kawai}}, \bibinfo {author} {\bibfnamefont {O.}~\bibnamefont {Neucheva}},
  \bibinfo {author} {\bibfnamefont {T.~L.}\ \bibnamefont {Yap}}, \bibinfo
  {author} {\bibfnamefont {C.}~\bibnamefont {Joachim}}, \ and\ \bibinfo
  {author} {\bibfnamefont {M.}~\bibnamefont {Saeys}},\ }\href@noop {}
  {\bibfield  {journal} {\bibinfo  {journal} {Surf. Sci.}\ }\textbf {\bibinfo
  {volume} {645}},\ \bibinfo {pages} {88} (\bibinfo {year} {2016})}\BibitemShut
  {NoStop}%
\bibitem [{\citenamefont {Scherpelz}\ and\ \citenamefont
  {Galli}(2017)}]{2017Scherpelz}%
  \BibitemOpen
  \bibfield  {author} {\bibinfo {author} {\bibfnamefont {P.}~\bibnamefont
  {Scherpelz}}\ and\ \bibinfo {author} {\bibfnamefont {G.}~\bibnamefont
  {Galli}},\ }\href@noop {} {\bibfield  {journal} {\bibinfo  {journal} {Phys.
  Rev. Materials}\ }\textbf {\bibinfo {volume} {1}},\ \bibinfo {pages} {021602}
  (\bibinfo {year} {2017})}\BibitemShut {NoStop}%
\bibitem [{\citenamefont {Lyo}\ and\ \citenamefont {Avouris}(1989)}]{1989Lyo}%
  \BibitemOpen
  \bibfield  {author} {\bibinfo {author} {\bibfnamefont {I.-W.}\ \bibnamefont
  {Lyo}}\ and\ \bibinfo {author} {\bibfnamefont {P.}~\bibnamefont {Avouris}},\
  }\href@noop {} {\bibfield  {journal} {\bibinfo  {journal} {Science}\ }\textbf
  {\bibinfo {volume} {245}},\ \bibinfo {pages} {1369} (\bibinfo {year}
  {1989})}\BibitemShut {NoStop}%
\bibitem [{\citenamefont {Schofield}\ \emph {et~al.}(2013)\citenamefont
  {Schofield}, \citenamefont {Studer}, \citenamefont {Hirjibehedin},
  \citenamefont {Curson}, \citenamefont {Aeppli},\ and\ \citenamefont
  {Bowler}}]{2013Schofield}%
  \BibitemOpen
  \bibfield  {author} {\bibinfo {author} {\bibfnamefont {S.~R.}\ \bibnamefont
  {Schofield}}, \bibinfo {author} {\bibfnamefont {P.}~\bibnamefont {Studer}},
  \bibinfo {author} {\bibfnamefont {C.~F.}\ \bibnamefont {Hirjibehedin}},
  \bibinfo {author} {\bibfnamefont {N.~J.}\ \bibnamefont {Curson}}, \bibinfo
  {author} {\bibfnamefont {G.}~\bibnamefont {Aeppli}}, \ and\ \bibinfo {author}
  {\bibfnamefont {D.~R.}\ \bibnamefont {Bowler}},\ }\href@noop {} {\bibfield
  {journal} {\bibinfo  {journal} {Nat. Commun.}\ }\textbf {\bibinfo {volume}
  {4}},\ \bibinfo {pages} {1649} (\bibinfo {year} {2013})}\BibitemShut
  {NoStop}%
\bibitem [{\citenamefont {Rashidi}\ \emph {et~al.}(2016)\citenamefont
  {Rashidi}, \citenamefont {Taucer}, \citenamefont {Ozfidan}, \citenamefont
  {Lloyd}, \citenamefont {Koleini}, \citenamefont {Labidi}, \citenamefont
  {Pitters}, \citenamefont {Maciejko},\ and\ \citenamefont
  {Wolkow}}]{2016Rashidi}%
  \BibitemOpen
  \bibfield  {author} {\bibinfo {author} {\bibfnamefont {M.}~\bibnamefont
  {Rashidi}}, \bibinfo {author} {\bibfnamefont {M.}~\bibnamefont {Taucer}},
  \bibinfo {author} {\bibfnamefont {I.}~\bibnamefont {Ozfidan}}, \bibinfo
  {author} {\bibfnamefont {E.}~\bibnamefont {Lloyd}}, \bibinfo {author}
  {\bibfnamefont {M.}~\bibnamefont {Koleini}}, \bibinfo {author} {\bibfnamefont
  {H.}~\bibnamefont {Labidi}}, \bibinfo {author} {\bibfnamefont {J.~L.}\
  \bibnamefont {Pitters}}, \bibinfo {author} {\bibfnamefont {J.}~\bibnamefont
  {Maciejko}}, \ and\ \bibinfo {author} {\bibfnamefont {R.~A.}\ \bibnamefont
  {Wolkow}},\ }\href@noop {} {\bibfield  {journal} {\bibinfo  {journal} {Phys.
  Rev. Lett.}\ }\textbf {\bibinfo {volume} {117}},\ \bibinfo {pages} {276805}
  (\bibinfo {year} {2016})}\BibitemShut {NoStop}%
\bibitem [{\citenamefont {Wyrick}\ \emph {et~al.}(2018)\citenamefont {Wyrick},
  \citenamefont {Wang}, \citenamefont {Namboodiri}, \citenamefont {Schmucker},
  \citenamefont {Kashid},\ and\ \citenamefont {Silver}}]{2018Wyrick}%
  \BibitemOpen
  \bibfield  {author} {\bibinfo {author} {\bibfnamefont {J.}~\bibnamefont
  {Wyrick}}, \bibinfo {author} {\bibfnamefont {X.}~\bibnamefont {Wang}},
  \bibinfo {author} {\bibfnamefont {P.}~\bibnamefont {Namboodiri}}, \bibinfo
  {author} {\bibfnamefont {S.~W.}\ \bibnamefont {Schmucker}}, \bibinfo {author}
  {\bibfnamefont {R.~V.}\ \bibnamefont {Kashid}}, \ and\ \bibinfo {author}
  {\bibfnamefont {R.~M.}\ \bibnamefont {Silver}},\ }\href@noop {} {\bibfield
  {journal} {\bibinfo  {journal} {Nano Lett.}\ }\textbf {\bibinfo {volume}
  {18}},\ \bibinfo {pages} {7502} (\bibinfo {year} {2018})}\BibitemShut
  {NoStop}%
\bibitem [{\citenamefont {Haider}\ \emph {et~al.}(2009)\citenamefont {Haider},
  \citenamefont {Pitters}, \citenamefont {DiLabio}, \citenamefont {Livadaru},
  \citenamefont {Mutus},\ and\ \citenamefont {Wolkow}}]{2009Haider}%
  \BibitemOpen
  \bibfield  {author} {\bibinfo {author} {\bibfnamefont {M.~B.}\ \bibnamefont
  {Haider}}, \bibinfo {author} {\bibfnamefont {J.~L.}\ \bibnamefont {Pitters}},
  \bibinfo {author} {\bibfnamefont {G.~A.}\ \bibnamefont {DiLabio}}, \bibinfo
  {author} {\bibfnamefont {L.}~\bibnamefont {Livadaru}}, \bibinfo {author}
  {\bibfnamefont {J.~Y.}\ \bibnamefont {Mutus}}, \ and\ \bibinfo {author}
  {\bibfnamefont {R.~A.}\ \bibnamefont {Wolkow}},\ }\href@noop {} {\bibfield
  {journal} {\bibinfo  {journal} {Phys. Rev. Lett.}\ }\textbf {\bibinfo
  {volume} {102}},\ \bibinfo {pages} {046805} (\bibinfo {year}
  {2009})}\BibitemShut {NoStop}%
\bibitem [{\citenamefont {He}\ \emph {et~al.}(2019)\citenamefont {He},
  \citenamefont {Gorman}, \citenamefont {Keith}, \citenamefont {Kranz},
  \citenamefont {Keizer},\ and\ \citenamefont {Simmons}}]{2019He}%
  \BibitemOpen
  \bibfield  {author} {\bibinfo {author} {\bibfnamefont {Y.}~\bibnamefont
  {He}}, \bibinfo {author} {\bibfnamefont {S.~K.}\ \bibnamefont {Gorman}},
  \bibinfo {author} {\bibfnamefont {D.}~\bibnamefont {Keith}}, \bibinfo
  {author} {\bibfnamefont {L.}~\bibnamefont {Kranz}}, \bibinfo {author}
  {\bibfnamefont {J.~G.}\ \bibnamefont {Keizer}}, \ and\ \bibinfo {author}
  {\bibfnamefont {M.~Y.}\ \bibnamefont {Simmons}},\ }\href@noop {} {\bibfield
  {journal} {\bibinfo  {journal} {Nature}\ }\textbf {\bibinfo {volume} {571}},\
  \bibinfo {pages} {371} (\bibinfo {year} {2019})}\BibitemShut {NoStop}%
\bibitem [{\citenamefont {Achal}\ \emph {et~al.}(2020)\citenamefont {Achal},
  \citenamefont {Rashidi}, \citenamefont {Croshaw}, \citenamefont {Huff},\ and\
  \citenamefont {Wolkow}}]{2020Achal}%
  \BibitemOpen
  \bibfield  {author} {\bibinfo {author} {\bibfnamefont {R.}~\bibnamefont
  {Achal}}, \bibinfo {author} {\bibfnamefont {M.}~\bibnamefont {Rashidi}},
  \bibinfo {author} {\bibfnamefont {J.}~\bibnamefont {Croshaw}}, \bibinfo
  {author} {\bibfnamefont {T.~R.}\ \bibnamefont {Huff}}, \ and\ \bibinfo
  {author} {\bibfnamefont {R.~A.}\ \bibnamefont {Wolkow}},\ }\href@noop {}
  {\bibfield  {journal} {\bibinfo  {journal} {ACS Nano}\ }\textbf {\bibinfo
  {volume} {14}},\ \bibinfo {pages} {2947} (\bibinfo {year}
  {2020})}\BibitemShut {NoStop}%
\bibitem [{\citenamefont {Stock}\ \emph {et~al.}(2020)\citenamefont {Stock},
  \citenamefont {Warschkow}, \citenamefont {Constantinou}, \citenamefont {Li},
  \citenamefont {Fearn}, \citenamefont {Crane}, \citenamefont {Hofmann},
  \citenamefont {Kölker}, \citenamefont {McKenzie}, \citenamefont
  {Schofield},\ and\ \citenamefont {Curson}}]{2020Stock}%
  \BibitemOpen
  \bibfield  {author} {\bibinfo {author} {\bibfnamefont {T.~J.~Z.}\
  \bibnamefont {Stock}}, \bibinfo {author} {\bibfnamefont {O.}~\bibnamefont
  {Warschkow}}, \bibinfo {author} {\bibfnamefont {P.~C.}\ \bibnamefont
  {Constantinou}}, \bibinfo {author} {\bibfnamefont {J.}~\bibnamefont {Li}},
  \bibinfo {author} {\bibfnamefont {S.}~\bibnamefont {Fearn}}, \bibinfo
  {author} {\bibfnamefont {E.}~\bibnamefont {Crane}}, \bibinfo {author}
  {\bibfnamefont {E.~V.~S.}\ \bibnamefont {Hofmann}}, \bibinfo {author}
  {\bibfnamefont {A.}~\bibnamefont {Kölker}}, \bibinfo {author} {\bibfnamefont
  {D.~R.}\ \bibnamefont {McKenzie}}, \bibinfo {author} {\bibfnamefont {S.~R.}\
  \bibnamefont {Schofield}}, \ and\ \bibinfo {author} {\bibfnamefont {N.~J.}\
  \bibnamefont {Curson}},\ }\href@noop {} {\bibfield  {journal} {\bibinfo
  {journal} {ACS Nano}\ }\textbf {\bibinfo {volume} {14}},\ \bibinfo {pages}
  {3316} (\bibinfo {year} {2020})}\BibitemShut {NoStop}%
\bibitem [{\citenamefont {Schofield}\ \emph {et~al.}(2003)\citenamefont
  {Schofield}, \citenamefont {Curson}, \citenamefont {Simmons}, \citenamefont
  {Rue\ss{}}, \citenamefont {Hallam}, \citenamefont {Oberbeck},\ and\
  \citenamefont {Clark}}]{2003Schofield}%
  \BibitemOpen
  \bibfield  {author} {\bibinfo {author} {\bibfnamefont {S.~R.}\ \bibnamefont
  {Schofield}}, \bibinfo {author} {\bibfnamefont {N.~J.}\ \bibnamefont
  {Curson}}, \bibinfo {author} {\bibfnamefont {M.~Y.}\ \bibnamefont {Simmons}},
  \bibinfo {author} {\bibfnamefont {F.~J.}\ \bibnamefont {Rue\ss{}}}, \bibinfo
  {author} {\bibfnamefont {T.}~\bibnamefont {Hallam}}, \bibinfo {author}
  {\bibfnamefont {L.}~\bibnamefont {Oberbeck}}, \ and\ \bibinfo {author}
  {\bibfnamefont {R.~G.}\ \bibnamefont {Clark}},\ }\href@noop {} {\bibfield
  {journal} {\bibinfo  {journal} {Phys. Rev. Lett.}\ }\textbf {\bibinfo
  {volume} {91}},\ \bibinfo {pages} {136104} (\bibinfo {year}
  {2003})}\BibitemShut {NoStop}%
\bibitem [{\citenamefont {Zikovsky}\ \emph {et~al.}(2011)\citenamefont
  {Zikovsky}, \citenamefont {Dogel}, \citenamefont {Salomons}, \citenamefont
  {Pitters}, \citenamefont {DiLabio},\ and\ \citenamefont
  {Wolkow}}]{2011Zikovsky}%
  \BibitemOpen
  \bibfield  {author} {\bibinfo {author} {\bibfnamefont {J.}~\bibnamefont
  {Zikovsky}}, \bibinfo {author} {\bibfnamefont {S.~A.}\ \bibnamefont {Dogel}},
  \bibinfo {author} {\bibfnamefont {M.~H.}\ \bibnamefont {Salomons}}, \bibinfo
  {author} {\bibfnamefont {J.~L.}\ \bibnamefont {Pitters}}, \bibinfo {author}
  {\bibfnamefont {G.~A.}\ \bibnamefont {DiLabio}}, \ and\ \bibinfo {author}
  {\bibfnamefont {R.~A.}\ \bibnamefont {Wolkow}},\ }\href@noop {} {\bibfield
  {journal} {\bibinfo  {journal} {J. Chem. Phys.}\ }\textbf {\bibinfo {volume}
  {134}},\ \bibinfo {pages} {114707} (\bibinfo {year} {2011})}\BibitemShut
  {NoStop}%
\bibitem [{\citenamefont {Li}\ \emph {et~al.}(2011)\citenamefont {Li},
  \citenamefont {Chang}, \citenamefont {Chien}, \citenamefont {Chang},
  \citenamefont {Chiang},\ and\ \citenamefont {Lin}}]{2011Li}%
  \BibitemOpen
  \bibfield  {author} {\bibinfo {author} {\bibfnamefont {H.-D.}\ \bibnamefont
  {Li}}, \bibinfo {author} {\bibfnamefont {C.-Y.}\ \bibnamefont {Chang}},
  \bibinfo {author} {\bibfnamefont {L.-Y.}\ \bibnamefont {Chien}}, \bibinfo
  {author} {\bibfnamefont {S.-H.}\ \bibnamefont {Chang}}, \bibinfo {author}
  {\bibfnamefont {T.-C.}\ \bibnamefont {Chiang}}, \ and\ \bibinfo {author}
  {\bibfnamefont {D.-S.}\ \bibnamefont {Lin}},\ }\href@noop {} {\bibfield
  {journal} {\bibinfo  {journal} {Phys. Rev. B}\ }\textbf {\bibinfo {volume}
  {83}},\ \bibinfo {pages} {075403} (\bibinfo {year} {2011})}\BibitemShut
  {NoStop}%
\bibitem [{\citenamefont {Ferng}\ and\ \citenamefont {Lin}(2012)}]{2012Ferng}%
  \BibitemOpen
  \bibfield  {author} {\bibinfo {author} {\bibfnamefont {S.-S.}\ \bibnamefont
  {Ferng}}\ and\ \bibinfo {author} {\bibfnamefont {D.-S.}\ \bibnamefont
  {Lin}},\ }\href@noop {} {\bibfield  {journal} {\bibinfo  {journal} {J. Phys.
  Chem. C}\ }\textbf {\bibinfo {volume} {116}},\ \bibinfo {pages} {3091}
  (\bibinfo {year} {2012})}\BibitemShut {NoStop}%
\bibitem [{\citenamefont {Mette}\ \emph {et~al.}(2009)\citenamefont {Mette},
  \citenamefont {Schwalb}, \citenamefont {Dürr},\ and\ \citenamefont
  {Höfer}}]{2009Mette}%
  \BibitemOpen
  \bibfield  {author} {\bibinfo {author} {\bibfnamefont {G.}~\bibnamefont
  {Mette}}, \bibinfo {author} {\bibfnamefont {C.}~\bibnamefont {Schwalb}},
  \bibinfo {author} {\bibfnamefont {M.}~\bibnamefont {Dürr}}, \ and\ \bibinfo
  {author} {\bibfnamefont {U.}~\bibnamefont {Höfer}},\ }\href@noop {}
  {\bibfield  {journal} {\bibinfo  {journal} {Chem. Phys. Lett.}\ }\textbf
  {\bibinfo {volume} {483}},\ \bibinfo {pages} {209} (\bibinfo {year}
  {2009})}\BibitemShut {NoStop}%
\bibitem [{\citenamefont {Cao}\ and\ \citenamefont {Hamers}(2002)}]{2002Cao}%
  \BibitemOpen
  \bibfield  {author} {\bibinfo {author} {\bibfnamefont {X.}~\bibnamefont
  {Cao}}\ and\ \bibinfo {author} {\bibfnamefont {R.~J.}\ \bibnamefont
  {Hamers}},\ }\href@noop {} {\bibfield  {journal} {\bibinfo  {journal} {J.
  Phys. Chem. B}\ }\textbf {\bibinfo {volume} {106}},\ \bibinfo {pages} {1840}
  (\bibinfo {year} {2002})}\BibitemShut {NoStop}%
\bibitem [{\citenamefont {Ryan}\ \emph {et~al.}(2012)\citenamefont {Ryan},
  \citenamefont {Livadaru}, \citenamefont {DiLabio},\ and\ \citenamefont
  {Wolkow}}]{2012Ryan}%
  \BibitemOpen
  \bibfield  {author} {\bibinfo {author} {\bibfnamefont {P.~M.}\ \bibnamefont
  {Ryan}}, \bibinfo {author} {\bibfnamefont {L.}~\bibnamefont {Livadaru}},
  \bibinfo {author} {\bibfnamefont {G.~A.}\ \bibnamefont {DiLabio}}, \ and\
  \bibinfo {author} {\bibfnamefont {R.~A.}\ \bibnamefont {Wolkow}},\
  }\href@noop {} {\bibfield  {journal} {\bibinfo  {journal} {J. Am. Chem.
  Soc.}\ }\textbf {\bibinfo {volume} {134}},\ \bibinfo {pages} {12054}
  (\bibinfo {year} {2012})}\BibitemShut {NoStop}%
\bibitem [{\citenamefont {Piva}\ \emph {et~al.}(2014)\citenamefont {Piva},
  \citenamefont {DiLabio}, \citenamefont {Livadaru},\ and\ \citenamefont
  {Wolkow}}]{2014Piva}%
  \BibitemOpen
  \bibfield  {author} {\bibinfo {author} {\bibfnamefont {P.~G.}\ \bibnamefont
  {Piva}}, \bibinfo {author} {\bibfnamefont {G.~A.}\ \bibnamefont {DiLabio}},
  \bibinfo {author} {\bibfnamefont {L.}~\bibnamefont {Livadaru}}, \ and\
  \bibinfo {author} {\bibfnamefont {R.~A.}\ \bibnamefont {Wolkow}},\
  }\href@noop {} {\bibfield  {journal} {\bibinfo  {journal} {Phys. Rev. B}\
  }\textbf {\bibinfo {volume} {90}},\ \bibinfo {pages} {155422} (\bibinfo
  {year} {2014})}\BibitemShut {NoStop}%
\bibitem [{\citenamefont {Pavlova}\ \emph {et~al.}(2018)\citenamefont
  {Pavlova}, \citenamefont {Zhidomirov},\ and\ \citenamefont
  {Eltsov}}]{2018Pavlova}%
  \BibitemOpen
  \bibfield  {author} {\bibinfo {author} {\bibfnamefont {T.~V.}\ \bibnamefont
  {Pavlova}}, \bibinfo {author} {\bibfnamefont {G.~M.}\ \bibnamefont
  {Zhidomirov}}, \ and\ \bibinfo {author} {\bibfnamefont {K.~N.}\ \bibnamefont
  {Eltsov}},\ }\href@noop {} {\bibfield  {journal} {\bibinfo  {journal} {J.
  Phys. Chem. C}\ }\textbf {\bibinfo {volume} {122}},\ \bibinfo {pages} {1741}
  (\bibinfo {year} {2018})}\BibitemShut {NoStop}%
\bibitem [{\citenamefont {Dwyer}\ \emph {et~al.}(2019)\citenamefont {Dwyer},
  \citenamefont {Dreyer},\ and\ \citenamefont {Butera}}]{2019Dwyer}%
  \BibitemOpen
  \bibfield  {author} {\bibinfo {author} {\bibfnamefont {K.~J.}\ \bibnamefont
  {Dwyer}}, \bibinfo {author} {\bibfnamefont {M.}~\bibnamefont {Dreyer}}, \
  and\ \bibinfo {author} {\bibfnamefont {R.~E.}\ \bibnamefont {Butera}},\
  }\href@noop {} {\bibfield  {journal} {\bibinfo  {journal} {J. Phys. Chem. A}\
  }\textbf {\bibinfo {volume} {123}},\ \bibinfo {pages} {10793} (\bibinfo
  {year} {2019})}\BibitemShut {NoStop}%
\bibitem [{\citenamefont {Pavlova}\ \emph
  {et~al.}(2020{\natexlab{a}})\citenamefont {Pavlova}, \citenamefont
  {Shevlyuga}, \citenamefont {Andryushechkin}, \citenamefont {Zhidomirov},\
  and\ \citenamefont {Eltsov}}]{2020Pavlova}%
  \BibitemOpen
  \bibfield  {author} {\bibinfo {author} {\bibfnamefont {T.~V.}\ \bibnamefont
  {Pavlova}}, \bibinfo {author} {\bibfnamefont {V.~M.}\ \bibnamefont
  {Shevlyuga}}, \bibinfo {author} {\bibfnamefont {B.~V.}\ \bibnamefont
  {Andryushechkin}}, \bibinfo {author} {\bibfnamefont {G.~M.}\ \bibnamefont
  {Zhidomirov}}, \ and\ \bibinfo {author} {\bibfnamefont {K.~N.}\ \bibnamefont
  {Eltsov}},\ }\href@noop {} {\bibfield  {journal} {\bibinfo  {journal} {Appl.
  Surf. Sci.}\ }\textbf {\bibinfo {volume} {509}},\ \bibinfo {pages} {145235}
  (\bibinfo {year} {2020}{\natexlab{a}})}\BibitemShut {NoStop}%
\bibitem [{\citenamefont {Pavlova}\ and\ \citenamefont
  {Eltsov}(2021)}]{2021Pavlova}%
  \BibitemOpen
  \bibfield  {author} {\bibinfo {author} {\bibfnamefont {T.~V.}\ \bibnamefont
  {Pavlova}}\ and\ \bibinfo {author} {\bibfnamefont {K.~N.}\ \bibnamefont
  {Eltsov}},\ }\href@noop {} {\bibfield  {journal} {\bibinfo  {journal} {J.
  Phys.: Condens. Matter}\ }\textbf {\bibinfo {volume} {33}},\ \bibinfo {pages}
  {384001} (\bibinfo {year} {2021})}\BibitemShut {NoStop}%
\bibitem [{\citenamefont {Frederick}\ \emph {et~al.}(2021)\citenamefont
  {Frederick}, \citenamefont {Dwyer}, \citenamefont {Wang}, \citenamefont
  {Misra},\ and\ \citenamefont {Butera}}]{2021Frederick}%
  \BibitemOpen
  \bibfield  {author} {\bibinfo {author} {\bibfnamefont {E.}~\bibnamefont
  {Frederick}}, \bibinfo {author} {\bibfnamefont {K.~J.}\ \bibnamefont
  {Dwyer}}, \bibinfo {author} {\bibfnamefont {G.~T.}\ \bibnamefont {Wang}},
  \bibinfo {author} {\bibfnamefont {S.}~\bibnamefont {Misra}}, \ and\ \bibinfo
  {author} {\bibfnamefont {R.~E.}\ \bibnamefont {Butera}},\ }\href@noop {}
  {\bibfield  {journal} {\bibinfo  {journal} {J. Phys.: Condens. Matter}\
  }\textbf {\bibinfo {volume} {33}},\ \bibinfo {pages} {444001} (\bibinfo
  {year} {2021})}\BibitemShut {NoStop}%
\bibitem [{\citenamefont {Boland}(1993)}]{1993Boland}%
  \BibitemOpen
  \bibfield  {author} {\bibinfo {author} {\bibfnamefont {J.~J.}\ \bibnamefont
  {Boland}},\ }\href@noop {} {\bibfield  {journal} {\bibinfo  {journal}
  {Science}\ }\textbf {\bibinfo {volume} {262}},\ \bibinfo {pages} {1703}
  (\bibinfo {year} {1993})}\BibitemShut {NoStop}%
\bibitem [{\citenamefont {Dohn\'{a}lek}\ \emph {et~al.}(1997)\citenamefont
  {Dohn\'{a}lek}, \citenamefont {Lyubinetsky},\ and\ \citenamefont
  {Yates}}]{1997Dohnalek}%
  \BibitemOpen
  \bibfield  {author} {\bibinfo {author} {\bibfnamefont {Z.}~\bibnamefont
  {Dohn\'{a}lek}}, \bibinfo {author} {\bibfnamefont {I.}~\bibnamefont
  {Lyubinetsky}}, \ and\ \bibinfo {author} {\bibfnamefont {J.~T.}\ \bibnamefont
  {Yates}},\ }\href@noop {} {\bibfield  {journal} {\bibinfo  {journal} {J. Vac.
  Sci. Technol. A}\ }\textbf {\bibinfo {volume} {15}},\ \bibinfo {pages} {1488}
  (\bibinfo {year} {1997})}\BibitemShut {NoStop}%
\bibitem [{\citenamefont {Lyubinetsky}\ \emph {et~al.}(1998)\citenamefont
  {Lyubinetsky}, \citenamefont {Dohn\'alek}, \citenamefont {Choyke},\ and\
  \citenamefont {Yates}}]{1998Lyubinetsky}%
  \BibitemOpen
  \bibfield  {author} {\bibinfo {author} {\bibfnamefont {I.}~\bibnamefont
  {Lyubinetsky}}, \bibinfo {author} {\bibfnamefont {Z.}~\bibnamefont
  {Dohn\'alek}}, \bibinfo {author} {\bibfnamefont {W.~J.}\ \bibnamefont
  {Choyke}}, \ and\ \bibinfo {author} {\bibfnamefont {J.~T.}\ \bibnamefont
  {Yates}},\ }\href@noop {} {\bibfield  {journal} {\bibinfo  {journal} {Phys.
  Rev. B}\ }\textbf {\bibinfo {volume} {58}},\ \bibinfo {pages} {7950}
  (\bibinfo {year} {1998})}\BibitemShut {NoStop}%
\bibitem [{\citenamefont {Nakayama}\ \emph {et~al.}(2002)\citenamefont
  {Nakayama}, \citenamefont {Graugnard},\ and\ \citenamefont
  {Weaver}}]{2002Nakayama}%
  \BibitemOpen
  \bibfield  {author} {\bibinfo {author} {\bibfnamefont {K.~S.}\ \bibnamefont
  {Nakayama}}, \bibinfo {author} {\bibfnamefont {E.}~\bibnamefont {Graugnard}},
  \ and\ \bibinfo {author} {\bibfnamefont {J.~H.}\ \bibnamefont {Weaver}},\
  }\href@noop {} {\bibfield  {journal} {\bibinfo  {journal} {Phys. Rev. Lett.}\
  }\textbf {\bibinfo {volume} {89}},\ \bibinfo {pages} {266106} (\bibinfo
  {year} {2002})}\BibitemShut {NoStop}%
\bibitem [{\citenamefont {Nakamura}\ \emph {et~al.}(2003)\citenamefont
  {Nakamura}, \citenamefont {Mera},\ and\ \citenamefont
  {Maeda}}]{2003Nakamura}%
  \BibitemOpen
  \bibfield  {author} {\bibinfo {author} {\bibfnamefont {Y.}~\bibnamefont
  {Nakamura}}, \bibinfo {author} {\bibfnamefont {Y.}~\bibnamefont {Mera}}, \
  and\ \bibinfo {author} {\bibfnamefont {K.}~\bibnamefont {Maeda}},\
  }\href@noop {} {\bibfield  {journal} {\bibinfo  {journal} {Surf. Sci.}\
  }\textbf {\bibinfo {volume} {531}},\ \bibinfo {pages} {68} (\bibinfo {year}
  {2003})}\BibitemShut {NoStop}%
\bibitem [{\citenamefont {Nakamura}\ \emph {et~al.}(2001)\citenamefont
  {Nakamura}, \citenamefont {Mera},\ and\ \citenamefont
  {Maeda}}]{2001Nakamura}%
  \BibitemOpen
  \bibfield  {author} {\bibinfo {author} {\bibfnamefont {Y.}~\bibnamefont
  {Nakamura}}, \bibinfo {author} {\bibfnamefont {Y.}~\bibnamefont {Mera}}, \
  and\ \bibinfo {author} {\bibfnamefont {K.}~\bibnamefont {Maeda}},\
  }\href@noop {} {\bibfield  {journal} {\bibinfo  {journal} {Surf. Sci.}\
  }\textbf {\bibinfo {volume} {487}},\ \bibinfo {pages} {127} (\bibinfo {year}
  {2001})}\BibitemShut {NoStop}%
\bibitem [{\citenamefont {Nakamura}\ \emph {et~al.}(2007)\citenamefont
  {Nakamura}, \citenamefont {Mera},\ and\ \citenamefont
  {Maeda}}]{2007Nakamura}%
  \BibitemOpen
  \bibfield  {author} {\bibinfo {author} {\bibfnamefont {Y.}~\bibnamefont
  {Nakamura}}, \bibinfo {author} {\bibfnamefont {Y.}~\bibnamefont {Mera}}, \
  and\ \bibinfo {author} {\bibfnamefont {K.}~\bibnamefont {Maeda}},\
  }\href@noop {} {\bibfield  {journal} {\bibinfo  {journal} {Surf. Sci.}\
  }\textbf {\bibinfo {volume} {601}},\ \bibinfo {pages} {2189} (\bibinfo {year}
  {2007})}\BibitemShut {NoStop}%
\bibitem [{\citenamefont {Mochiji}\ and\ \citenamefont
  {Ichikawa}(1999)}]{1999Mochiji}%
  \BibitemOpen
  \bibfield  {author} {\bibinfo {author} {\bibfnamefont {K.}~\bibnamefont
  {Mochiji}}\ and\ \bibinfo {author} {\bibfnamefont {M.}~\bibnamefont
  {Ichikawa}},\ }\href@noop {} {\bibfield  {journal} {\bibinfo  {journal} {Jpn.
  J. Appl. Phys.}\ }\textbf {\bibinfo {volume} {38}},\ \bibinfo {pages} {L1}
  (\bibinfo {year} {1999})}\BibitemShut {NoStop}%
\bibitem [{\citenamefont {Mochiji}\ and\ \citenamefont
  {Ichikawa}(2000)}]{2000Mochiji}%
  \BibitemOpen
  \bibfield  {author} {\bibinfo {author} {\bibfnamefont {K.}~\bibnamefont
  {Mochiji}}\ and\ \bibinfo {author} {\bibfnamefont {M.}~\bibnamefont
  {Ichikawa}},\ }\href@noop {} {\bibfield  {journal} {\bibinfo  {journal}
  {Phys. Rev. B}\ }\textbf {\bibinfo {volume} {62}},\ \bibinfo {pages} {2029}
  (\bibinfo {year} {2000})}\BibitemShut {NoStop}%
\bibitem [{\citenamefont {Lee}\ and\ \citenamefont {Kang}(2004)}]{2004Lee}%
  \BibitemOpen
  \bibfield  {author} {\bibinfo {author} {\bibfnamefont {J.~Y.}\ \bibnamefont
  {Lee}}\ and\ \bibinfo {author} {\bibfnamefont {M.-H.}\ \bibnamefont {Kang}},\
  }\href@noop {} {\bibfield  {journal} {\bibinfo  {journal} {Phys. Rev. B}\
  }\textbf {\bibinfo {volume} {69}},\ \bibinfo {pages} {113307} (\bibinfo
  {year} {2004})}\BibitemShut {NoStop}%
\bibitem [{\citenamefont {Horcas}\ \emph {et~al.}(2007)\citenamefont {Horcas},
  \citenamefont {Fernández}, \citenamefont {Gómez-Rodríguez}, \citenamefont
  {Colchero}, \citenamefont {Gómez-Herrero},\ and\ \citenamefont
  {Baro}}]{WSXM}%
  \BibitemOpen
  \bibfield  {author} {\bibinfo {author} {\bibfnamefont {I.}~\bibnamefont
  {Horcas}}, \bibinfo {author} {\bibfnamefont {R.}~\bibnamefont {Fernández}},
  \bibinfo {author} {\bibfnamefont {J.~M.}\ \bibnamefont {Gómez-Rodríguez}},
  \bibinfo {author} {\bibfnamefont {J.}~\bibnamefont {Colchero}}, \bibinfo
  {author} {\bibfnamefont {J.}~\bibnamefont {Gómez-Herrero}}, \ and\ \bibinfo
  {author} {\bibfnamefont {A.~M.}\ \bibnamefont {Baro}},\ }\href@noop {}
  {\bibfield  {journal} {\bibinfo  {journal} {Rev. Sci. Instrum.}\ }\textbf
  {\bibinfo {volume} {78}},\ \bibinfo {pages} {013705} (\bibinfo {year}
  {2007})}\BibitemShut {NoStop}%
\bibitem [{\citenamefont {Kresse}\ and\ \citenamefont
  {Furthm\"uller}(1996)}]{1996Kresse}%
  \BibitemOpen
  \bibfield  {author} {\bibinfo {author} {\bibfnamefont {G.}~\bibnamefont
  {Kresse}}\ and\ \bibinfo {author} {\bibfnamefont {J.}~\bibnamefont
  {Furthm\"uller}},\ }\href@noop {} {\bibfield  {journal} {\bibinfo  {journal}
  {Phys. Rev. B}\ }\textbf {\bibinfo {volume} {54}},\ \bibinfo {pages} {11169}
  (\bibinfo {year} {1996})}\BibitemShut {NoStop}%
\bibitem [{\citenamefont {Kresse}\ and\ \citenamefont
  {Joubert}(1999)}]{1999Kresse}%
  \BibitemOpen
  \bibfield  {author} {\bibinfo {author} {\bibfnamefont {G.}~\bibnamefont
  {Kresse}}\ and\ \bibinfo {author} {\bibfnamefont {D.}~\bibnamefont
  {Joubert}},\ }\href@noop {} {\bibfield  {journal} {\bibinfo  {journal} {Phys.
  Rev. B}\ }\textbf {\bibinfo {volume} {59}},\ \bibinfo {pages} {1758}
  (\bibinfo {year} {1999})}\BibitemShut {NoStop}%
\bibitem [{\citenamefont {Perdew}\ \emph {et~al.}(1996)\citenamefont {Perdew},
  \citenamefont {Burke},\ and\ \citenamefont {Ernzerhof}}]{1996Perdew}%
  \BibitemOpen
  \bibfield  {author} {\bibinfo {author} {\bibfnamefont {J.~P.}\ \bibnamefont
  {Perdew}}, \bibinfo {author} {\bibfnamefont {K.}~\bibnamefont {Burke}}, \
  and\ \bibinfo {author} {\bibfnamefont {M.}~\bibnamefont {Ernzerhof}},\
  }\href@noop {} {\bibfield  {journal} {\bibinfo  {journal} {Phys. Rev. Lett.}\
  }\textbf {\bibinfo {volume} {77}},\ \bibinfo {pages} {3865} (\bibinfo {year}
  {1996})}\BibitemShut {NoStop}%
\bibitem [{\citenamefont {Grimme}(2006)}]{2006Grimme}%
  \BibitemOpen
  \bibfield  {author} {\bibinfo {author} {\bibfnamefont {S.}~\bibnamefont
  {Grimme}},\ }\href@noop {} {\bibfield  {journal} {\bibinfo  {journal} {J.
  Comput. Chem.}\ }\textbf {\bibinfo {volume} {27}},\ \bibinfo {pages} {1787}
  (\bibinfo {year} {2006})}\BibitemShut {NoStop}%
\bibitem [{\citenamefont {Tersoff}\ and\ \citenamefont
  {Hamann}(1985)}]{1985Tersoff}%
  \BibitemOpen
  \bibfield  {author} {\bibinfo {author} {\bibfnamefont {J.}~\bibnamefont
  {Tersoff}}\ and\ \bibinfo {author} {\bibfnamefont {D.~R.}\ \bibnamefont
  {Hamann}},\ }\href@noop {} {\bibfield  {journal} {\bibinfo  {journal} {Phys.
  Rev. B}\ }\textbf {\bibinfo {volume} {31}},\ \bibinfo {pages} {805} (\bibinfo
  {year} {1985})}\BibitemShut {NoStop}%
\bibitem [{\citenamefont {M{\o}ller}\ \emph {et~al.}(2017)\citenamefont
  {M{\o}ller}, \citenamefont {Jarvis}, \citenamefont {Gu\'{e}rinet},
  \citenamefont {Sharp}, \citenamefont {Woolley}, \citenamefont {Rahe},\ and\
  \citenamefont {Moriarty}}]{2017Moller}%
  \BibitemOpen
  \bibfield  {author} {\bibinfo {author} {\bibfnamefont {M.}~\bibnamefont
  {M{\o}ller}}, \bibinfo {author} {\bibfnamefont {S.~P.}\ \bibnamefont
  {Jarvis}}, \bibinfo {author} {\bibfnamefont {L.}~\bibnamefont
  {Gu\'{e}rinet}}, \bibinfo {author} {\bibfnamefont {P.}~\bibnamefont {Sharp}},
  \bibinfo {author} {\bibfnamefont {R.}~\bibnamefont {Woolley}}, \bibinfo
  {author} {\bibfnamefont {P.}~\bibnamefont {Rahe}}, \ and\ \bibinfo {author}
  {\bibfnamefont {P.}~\bibnamefont {Moriarty}},\ }\href@noop {} {\bibfield
  {journal} {\bibinfo  {journal} {Nanotechnology}\ }\textbf {\bibinfo {volume}
  {28}},\ \bibinfo {pages} {075302} (\bibinfo {year} {2017})}\BibitemShut
  {NoStop}%
\bibitem [{\citenamefont {Pavlova}\ \emph
  {et~al.}(2020{\natexlab{b}})\citenamefont {Pavlova}, \citenamefont
  {Shevlyuga}, \citenamefont {Andryushechkin},\ and\ \citenamefont
  {Eltsov}}]{2020PavlovaPRB}%
  \BibitemOpen
  \bibfield  {author} {\bibinfo {author} {\bibfnamefont {T.~V.}\ \bibnamefont
  {Pavlova}}, \bibinfo {author} {\bibfnamefont {V.~M.}\ \bibnamefont
  {Shevlyuga}}, \bibinfo {author} {\bibfnamefont {B.~V.}\ \bibnamefont
  {Andryushechkin}}, \ and\ \bibinfo {author} {\bibfnamefont {K.~N.}\
  \bibnamefont {Eltsov}},\ }\href@noop {} {\bibfield  {journal} {\bibinfo
  {journal} {Phys. Rev. B}\ }\textbf {\bibinfo {volume} {101}},\ \bibinfo
  {pages} {235410} (\bibinfo {year} {2020}{\natexlab{b}})}\BibitemShut
  {NoStop}%
\bibitem [{\citenamefont {Ballard}\ \emph {et~al.}(2014)\citenamefont
  {Ballard}, \citenamefont {Owen}, \citenamefont {Alexander}, \citenamefont
  {Owen}, \citenamefont {Fuchs}, \citenamefont {Randall}, \citenamefont
  {Longo},\ and\ \citenamefont {Cho}}]{2014Ballard2}%
  \BibitemOpen
  \bibfield  {author} {\bibinfo {author} {\bibfnamefont {J.~B.}\ \bibnamefont
  {Ballard}}, \bibinfo {author} {\bibfnamefont {J.~H.~G.}\ \bibnamefont
  {Owen}}, \bibinfo {author} {\bibfnamefont {J.~D.}\ \bibnamefont {Alexander}},
  \bibinfo {author} {\bibfnamefont {W.~R.}\ \bibnamefont {Owen}}, \bibinfo
  {author} {\bibfnamefont {E.}~\bibnamefont {Fuchs}}, \bibinfo {author}
  {\bibfnamefont {J.~N.}\ \bibnamefont {Randall}}, \bibinfo {author}
  {\bibfnamefont {R.~C.}\ \bibnamefont {Longo}}, \ and\ \bibinfo {author}
  {\bibfnamefont {K.}~\bibnamefont {Cho}},\ }\href@noop {} {\bibfield
  {journal} {\bibinfo  {journal} {J. Vac. Sci. Technol. B}\ }\textbf {\bibinfo
  {volume} {32}},\ \bibinfo {pages} {021805} (\bibinfo {year}
  {2014})}\BibitemShut {NoStop}%
\bibitem [{\citenamefont {Pavlova}(2020)}]{2020PavlovaPCCP}%
  \BibitemOpen
  \bibfield  {author} {\bibinfo {author} {\bibfnamefont {T.~V.}\ \bibnamefont
  {Pavlova}},\ }\href@noop {} {\bibfield  {journal} {\bibinfo  {journal} {Phys.
  Chem. Chem. Phys.}\ }\textbf {\bibinfo {volume} {22}},\ \bibinfo {pages}
  {21851} (\bibinfo {year} {2020})}\BibitemShut {NoStop}%
\bibitem [{\citenamefont {Achal}\ \emph {et~al.}(2018)\citenamefont {Achal},
  \citenamefont {Rashidi}, \citenamefont {Croshaw}, \citenamefont {Churchill},
  \citenamefont {Taucer}, \citenamefont {Huff}, \citenamefont {Cloutier},
  \citenamefont {Pitters},\ and\ \citenamefont {Wolkow}}]{2018Achal}%
  \BibitemOpen
  \bibfield  {author} {\bibinfo {author} {\bibfnamefont {R.}~\bibnamefont
  {Achal}}, \bibinfo {author} {\bibfnamefont {M.}~\bibnamefont {Rashidi}},
  \bibinfo {author} {\bibfnamefont {J.}~\bibnamefont {Croshaw}}, \bibinfo
  {author} {\bibfnamefont {D.}~\bibnamefont {Churchill}}, \bibinfo {author}
  {\bibfnamefont {M.}~\bibnamefont {Taucer}}, \bibinfo {author} {\bibfnamefont
  {T.}~\bibnamefont {Huff}}, \bibinfo {author} {\bibfnamefont {M.}~\bibnamefont
  {Cloutier}}, \bibinfo {author} {\bibfnamefont {J.}~\bibnamefont {Pitters}}, \
  and\ \bibinfo {author} {\bibfnamefont {R.~A.}\ \bibnamefont {Wolkow}},\
  }\href@noop {} {\bibfield  {journal} {\bibinfo  {journal} {Nat. Comm.}\
  }\textbf {\bibinfo {volume} {9}},\ \bibinfo {pages} {2778} (\bibinfo {year}
  {2018})}\BibitemShut {NoStop}%
\bibitem [{\citenamefont {Randall}\ \emph {et~al.}(2018)\citenamefont
  {Randall}, \citenamefont {Owen}, \citenamefont {Fuchs}, \citenamefont {Lake},
  \citenamefont {Ehr}, \citenamefont {Ballard},\ and\ \citenamefont
  {Henriksen}}]{2018Randall}%
  \BibitemOpen
  \bibfield  {author} {\bibinfo {author} {\bibfnamefont {J.~N.}\ \bibnamefont
  {Randall}}, \bibinfo {author} {\bibfnamefont {J.~H.}\ \bibnamefont {Owen}},
  \bibinfo {author} {\bibfnamefont {E.}~\bibnamefont {Fuchs}}, \bibinfo
  {author} {\bibfnamefont {J.}~\bibnamefont {Lake}}, \bibinfo {author}
  {\bibfnamefont {J.~R.~V.}\ \bibnamefont {Ehr}}, \bibinfo {author}
  {\bibfnamefont {J.}~\bibnamefont {Ballard}}, \ and\ \bibinfo {author}
  {\bibfnamefont {E.}~\bibnamefont {Henriksen}},\ }\href@noop {} {\bibfield
  {journal} {\bibinfo  {journal} {Micro and Nano Engineering}\ }\textbf
  {\bibinfo {volume} {1}},\ \bibinfo {pages} {1} (\bibinfo {year}
  {2018})}\BibitemShut {NoStop}%
\bibitem [{\citenamefont {Shen}\ \emph {et~al.}(1995)\citenamefont {Shen},
  \citenamefont {Wang}, \citenamefont {Abeln}, \citenamefont {Tucker},
  \citenamefont {Lyding}, \citenamefont {Avouris},\ and\ \citenamefont
  {Walkup}}]{1995Shen}%
  \BibitemOpen
  \bibfield  {author} {\bibinfo {author} {\bibfnamefont {T.~C.}\ \bibnamefont
  {Shen}}, \bibinfo {author} {\bibfnamefont {C.}~\bibnamefont {Wang}}, \bibinfo
  {author} {\bibfnamefont {G.~C.}\ \bibnamefont {Abeln}}, \bibinfo {author}
  {\bibfnamefont {J.~R.}\ \bibnamefont {Tucker}}, \bibinfo {author}
  {\bibfnamefont {J.~W.}\ \bibnamefont {Lyding}}, \bibinfo {author}
  {\bibfnamefont {P.}~\bibnamefont {Avouris}}, \ and\ \bibinfo {author}
  {\bibfnamefont {R.~E.}\ \bibnamefont {Walkup}},\ }\href@noop {} {\bibfield
  {journal} {\bibinfo  {journal} {Science}\ }\textbf {\bibinfo {volume}
  {268}},\ \bibinfo {pages} {1590} (\bibinfo {year} {1995})}\BibitemShut
  {NoStop}%
\bibitem [{\citenamefont {Bellec}\ \emph {et~al.}(2010)\citenamefont {Bellec},
  \citenamefont {Riedel}, \citenamefont {Dujardin}, \citenamefont {Boudrioua},
  \citenamefont {Chaput}, \citenamefont {Stauffer},\ and\ \citenamefont
  {Sonnet}}]{2010Bellec}%
  \BibitemOpen
  \bibfield  {author} {\bibinfo {author} {\bibfnamefont {A.}~\bibnamefont
  {Bellec}}, \bibinfo {author} {\bibfnamefont {D.}~\bibnamefont {Riedel}},
  \bibinfo {author} {\bibfnamefont {G.}~\bibnamefont {Dujardin}}, \bibinfo
  {author} {\bibfnamefont {O.}~\bibnamefont {Boudrioua}}, \bibinfo {author}
  {\bibfnamefont {L.}~\bibnamefont {Chaput}}, \bibinfo {author} {\bibfnamefont
  {L.}~\bibnamefont {Stauffer}}, \ and\ \bibinfo {author} {\bibfnamefont
  {P.}~\bibnamefont {Sonnet}},\ }\href@noop {} {\bibfield  {journal} {\bibinfo
  {journal} {Phys. Rev. Lett.}\ }\textbf {\bibinfo {volume} {105}},\ \bibinfo
  {pages} {048302} (\bibinfo {year} {2010})}\BibitemShut {NoStop}%
\bibitem [{\citenamefont {Tong}\ and\ \citenamefont {Wolkow}(2006)}]{2006Tong}%
  \BibitemOpen
  \bibfield  {author} {\bibinfo {author} {\bibfnamefont {X.}~\bibnamefont
  {Tong}}\ and\ \bibinfo {author} {\bibfnamefont {R.~A.}\ \bibnamefont
  {Wolkow}},\ }\href@noop {} {\bibfield  {journal} {\bibinfo  {journal} {Surf.
  Sci.}\ }\textbf {\bibinfo {volume} {600}},\ \bibinfo {pages} {L199} (\bibinfo
  {year} {2006})}\BibitemShut {NoStop}%
\bibitem [{\citenamefont {Pavliček}\ \emph {et~al.}(2017)\citenamefont
  {Pavliček}, \citenamefont {Majzik}, \citenamefont {Meyer},\ and\
  \citenamefont {Gross}}]{2017Niko}%
  \BibitemOpen
  \bibfield  {author} {\bibinfo {author} {\bibfnamefont {N.}~\bibnamefont
  {Pavliček}}, \bibinfo {author} {\bibfnamefont {Z.}~\bibnamefont {Majzik}},
  \bibinfo {author} {\bibfnamefont {G.}~\bibnamefont {Meyer}}, \ and\ \bibinfo
  {author} {\bibfnamefont {L.}~\bibnamefont {Gross}},\ }\href@noop {}
  {\bibfield  {journal} {\bibinfo  {journal} {Appl. Phys. Lett.}\ }\textbf
  {\bibinfo {volume} {111}},\ \bibinfo {pages} {053104} (\bibinfo {year}
  {2017})}\BibitemShut {NoStop}%
\bibitem [{\citenamefont {Blomquist}\ and\ \citenamefont
  {Kirczenow}(2006)}]{2006Blomquist}%
  \BibitemOpen
  \bibfield  {author} {\bibinfo {author} {\bibfnamefont {T.}~\bibnamefont
  {Blomquist}}\ and\ \bibinfo {author} {\bibfnamefont {G.}~\bibnamefont
  {Kirczenow}},\ }\href@noop {} {\bibfield  {journal} {\bibinfo  {journal}
  {Phys. Rev. B}\ }\textbf {\bibinfo {volume} {73}},\ \bibinfo {pages} {195303}
  (\bibinfo {year} {2006})}\BibitemShut {NoStop}%
\bibitem [{\citenamefont {Lopinski}\ \emph {et~al.}(2005)\citenamefont
  {Lopinski}, \citenamefont {Eves}, \citenamefont {Hul'ko}, \citenamefont
  {Mark}, \citenamefont {Patitsas}, \citenamefont {Boukherroub},\ and\
  \citenamefont {Ward}}]{2005Lopinski}%
  \BibitemOpen
  \bibfield  {author} {\bibinfo {author} {\bibfnamefont {G.~P.}\ \bibnamefont
  {Lopinski}}, \bibinfo {author} {\bibfnamefont {B.~J.}\ \bibnamefont {Eves}},
  \bibinfo {author} {\bibfnamefont {O.}~\bibnamefont {Hul'ko}}, \bibinfo
  {author} {\bibfnamefont {C.}~\bibnamefont {Mark}}, \bibinfo {author}
  {\bibfnamefont {S.~N.}\ \bibnamefont {Patitsas}}, \bibinfo {author}
  {\bibfnamefont {R.}~\bibnamefont {Boukherroub}}, \ and\ \bibinfo {author}
  {\bibfnamefont {T.~R.}\ \bibnamefont {Ward}},\ }\href@noop {} {\bibfield
  {journal} {\bibinfo  {journal} {Phys. Rev. B}\ }\textbf {\bibinfo {volume}
  {71}},\ \bibinfo {pages} {125308} (\bibinfo {year} {2005})}\BibitemShut
  {NoStop}%
\bibitem [{\citenamefont {Pitters}\ \emph {et~al.}(2012)\citenamefont
  {Pitters}, \citenamefont {Piva},\ and\ \citenamefont {Wolkow}}]{2012Pitters}%
  \BibitemOpen
  \bibfield  {author} {\bibinfo {author} {\bibfnamefont {J.~L.}\ \bibnamefont
  {Pitters}}, \bibinfo {author} {\bibfnamefont {P.~G.}\ \bibnamefont {Piva}}, \
  and\ \bibinfo {author} {\bibfnamefont {R.~A.}\ \bibnamefont {Wolkow}},\
  }\href@noop {} {\bibfield  {journal} {\bibinfo  {journal} {J. Vac. Sci. Tech.
  B}\ }\textbf {\bibinfo {volume} {30}},\ \bibinfo {pages} {021806} (\bibinfo
  {year} {2012})}\BibitemShut {NoStop}%
\end{thebibliography}%

\end{document}